\documentclass[aps,prb]{revtex4}
\usepackage{epsfig}
\usepackage{color}
\usepackage{amsmath}
\usepackage{amssymb}
\usepackage{bbm}

\newcommand{\be}{\begin{equation}}
\newcommand{\ee}{\end{equation}}
\newcommand{\bea}{\begin{eqnarray}}
\newcommand{\eea}{\end{eqnarray}}

\begin{document}

\title{Overview of superconductivity in field-cooled magnetic materials.}

\author{Naoum Karchev}

\affiliation{Department of Physics, University of Sofia, 1164 Sofia, Bulgaria}

\begin{abstract}
	
	Considerable experimental skills have been accumulated in the preparation of field-cooled (FC) magnetic materials. This stimulates the search for FC magnetic materials that are superconductors.
	
	The article overviews the recent proposed mechanism of superconductivity in field-cooled  magnetic materials.
	It is based on previously published results for magnon-induced superconductivity in field-cooled spin-1/2 antiferromagnets $[PRB96,214409]$  (arXiv:1712.02983), and Sequence of superconducting states in field cooled $FeCr_2S_4$ $[JPCM33,495604]$  (arXiv:2111.02765).
	
	Shortened version of arXiv:2308.00470.

\end{abstract}

\maketitle

\section {INTRODUCTION}

The study of the mechanisms of superconductivity and the methods to synthesize superconductors is an exciting challenge in solid state physics.
Since the discovery of superconductivity (1911), these two activities have been a major driver for the successful dissemination of superconductivity ideas in other fields of physic and for  application in technology.

One way to prepare superconducting state  is to subject a suitable material to strong hydrostatic pressure. The coexistence of  ferromagnetism and superconductivity in $UGe_2$ under pressure was reported in the paper \cite{UGe2000}. The invention triggered a very intense experimental and theoretical study of the phenomenon \cite{Huxley2001,Tateiwa2001b,Motoyama2001,Huxley2002,Pfleiderer2009,Aoki2019}. 

The theoretical predictions are important, since they allow to focus experimental research on specific chemical compositions and also to suggest appropriate conditions for the synthesis.	
In his theoretical studies Ashcroft predicted that metalized hydrogen \cite{Ashcroft68} or hydrogen rich alloys \cite{Ashcroft04} can possess high temperature superconductivity. Ashcroft's idea is that the hydrogen is the lightest element, and if it can be compressed in a solid state, it could become superconductor at a very high transition temperature $T_c$ due to the strong electron-phonon coupling.
Accurate electronic structure and electron-phonon coupling calculations predicted high $T_c$ for metallic hydrogen \cite{Gross10,McMahon12,Bernstein15,Duan15,Pickard15}. Thanks the development of high-pressure techniques numerous experiments 
discussed the prediction. An important discovery leading to room-temperature superconductivity is the pressure-driven hydrogen sulfide with a confirmed transition temperature of 203 K at 155 GPa \cite{Drozdov15}. The most recent examples of a metal hydride are lanthanum hydride which has $Tc = 250 - 260 K$ at $ 180 - 200 GPa $ \cite{Drozdov19,Somayazulu19,Hong20} and carbonaceous sulfur hydride with room-temperature $T_c =287.7 K$ achieved at $267GPa$ \cite{Snider20}. More than ten hydrogen-rich compounds under high pressure have been found to be high temperature conventional superconductors: yttrium superhydride \cite{Troyan21,Snider21}, thorium hydride \cite{Semenok20}, praseodymium superhydride \cite{Zhou20}, barium superhydride \cite{Chen21} and others \cite{Sanna16,Gorkov18,Meng19,RevHSc20,Pickard20}.

Another way to fabricate unconventional superconductor is by chemical manipulation. The most famous example is copper-oxide superconductor.  The parent compound $La_2CuO_4$ is  an antiferromagnetic Mott insulator with N\'{e}el  temperature  $T_N = 300 K$.  The parent compound can be doped by substituting some of the trivalent $La$ by divalent $Sr$. The result is that
$x$ holes are added to the $Cu-O$ plane in $La_{2-x}Sr_xCuO_4$, which is called hole doping. The hole-doping suppresses  the antiferromagnetic order and at $x=0.03-0.05$ hole concentration
the system undergoes quantum antiferromagnetic-paramagnetic transition. After suppression of the antiferromagnetism, superconductivity appears, ranging from $x=0.06-0.25$ \cite{Bednorz86}. 
The electron doping is realized in the compound  $Nd_{2-x}Ce_xCuO_4$ \cite{Tokura89} when $x$ electrons are added. Details are given in many review articles and books, for example \cite{Ginsberg89,Dagotto94,Anderson97,Lee06}.

In the present Overview we discuss theoretically the emergence of superconductivity in field cooled magnetic materials. As examples we consider spin $s=1/2$ antiferromagnetic insulator and chromium spinel $FeCr_2S_4$. We also discuss some perspective antiferromagnetic compounds. 
 
\section {FIELD COOLED MAGNETIC MATERIALS}

\begin{figure}[!ht]
\centering\includegraphics[width=5in]{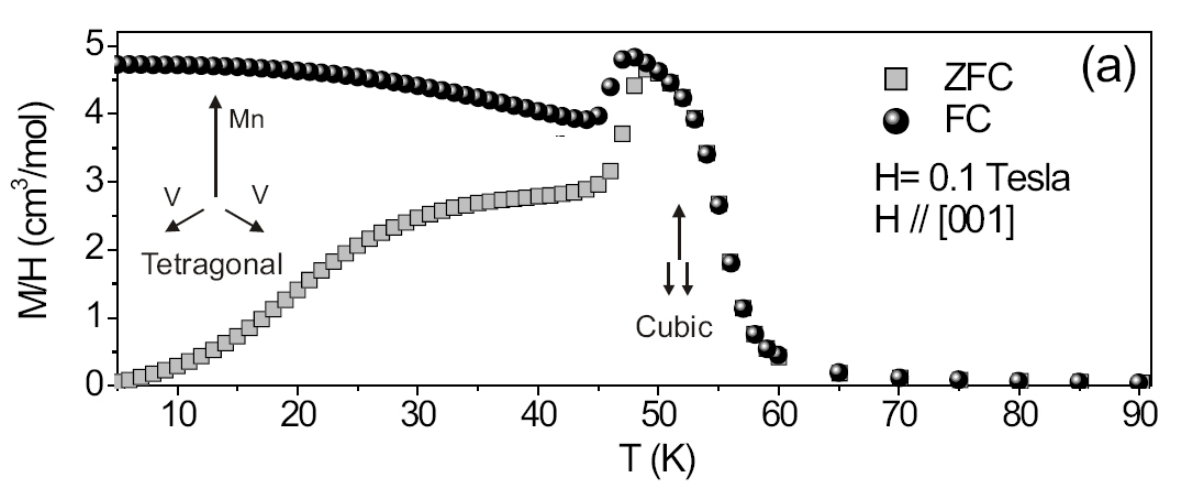}
\epsfxsize=\linewidth
\caption{Magnetization of ZFC and FC vanadium spinel as a function of temperature
	(after V.O.Garlea et al. PRL 100, 066404 (2008))}\label{(fig1)MnVexp}
\end{figure}

The material is field cooled (FC) if, during the preparation, an external magnetic field as high as $300 Oe$ is applied upon cooling. If the applied field is below $1 Oe$ it is zero field cooled (ZFC). The magnetization-temperature and magnetic susceptibility curves for (ZFC) and (FC) spinel show a remarkable difference below N\'{e}el $T_N$ temperature 
\cite{spinel+,spinelFeCr2S4,spinelCv1,spinel08,spinel++,spinel11b,spinel11a,spinel11c,spinel12a,spinel12b,spinel+1,spinel+2,spinel+3}. In the case of vanadium spinel $MnV_2O_4$ the curves, which show the temperature dependence of spontaneous magnetization $M$, are depicted in Fig.\ref{(fig1)MnVexp}. 

\begin{figure}[!ht]
\centering\includegraphics[width=4 in]{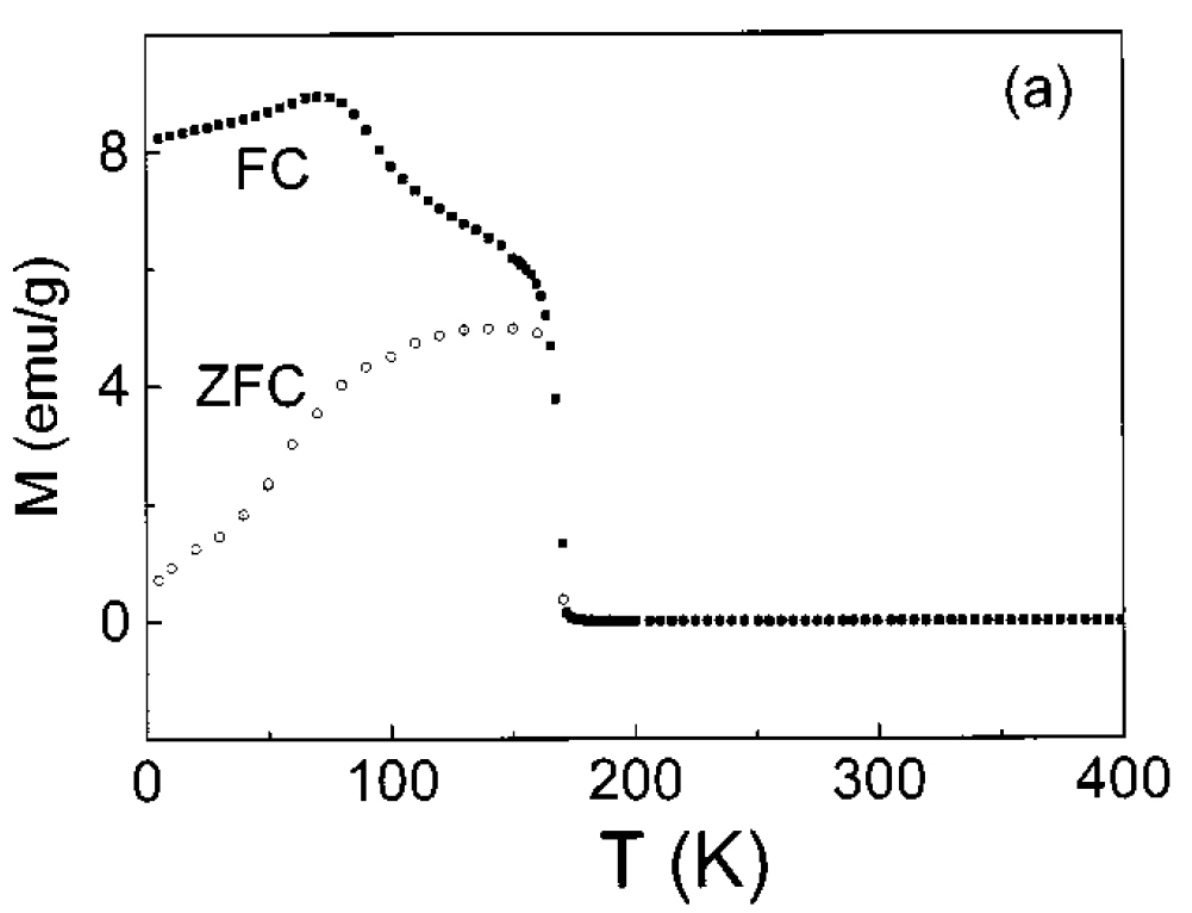}
	\epsfxsize=\linewidth
\caption{Magnetization of ZFC and FC chromium spinel as a function of temperature
		(after Zhaorong Yang at al. Phys Rev. {\bf B 62}, 13872 (2000))}\label{(fig2)FeCr2S4}
\end{figure}

The spinel $MnV_2O_4$ is a two-sublattices ferrimagnet, with site $A$ occupied by the $Mn^{2+}$ ion, which is in the $3d^5$ high-spin configuration with quenched orbital angular momentum, that can be regarded as a simple $s=5/2$ spin. The B site is occupied by the $V^{3+}$ ion, which takes the $3d^{2}$ high-spin configuration in the triply degenerate $t_{2g}$ orbital, and has orbital degrees of freedom. The measurements show that the set in of the magnetic order is at Neel temperature $T_N=56.5K$ \cite{spinel+}, and that the magnetization has a maximum near $T^*=53.5K$. Below this temperature the magnetization sharply decreases and goes to zero when temperature approaches zero.  The ferrimagnetic
phase of vanadium spinel is divided into two phases: high temperature  $(T^*,T_N)$, where the magnetization-temperature curves of (ZFC) and (FC) materials coincide, and low temperature  $0<T<T^*$, where there is a pronounced difference between the two curves.
The two phases are basic characteristic  of field cooled magnetic materials. As a second example we consider chromium spinel $FeCr_2S_4$. The magnetization-temperature curves are depicted in Fig.\ref{(fig2)FeCr2S4}.
\vskip 0.5 cm
{\bf PARTIAL ORDER AND PARTIAL ORDER TRANSITION}
\vskip 0.1 cm
  Magnetic state is a partial order state if only part of the electrons in the system give contribution to the magnetic order. It is studied in exactly solvable models \cite{Larkin66,Diep87}, by means of Green's function approach \cite{Diep97}, Monte Carlo method \cite{Diep87} and modified spin-wave theory of magnetism \cite{Karchev15}. 
  
  The spinel is typical example of a system with partial order and partial order transition. The vanadium spinel is a system that obtains its magnetic properties from $Mn$ and $V$ magnetic moments. The true magnons in this system, which are the transversal fluctuations corresponding to the total magnetization, are complicated mixtures of the $Mn$ and $V$ transversal fluctuations. The magnons interact with manganese and vanadium ions in a different way, and the magnons fluctuations
  suppress the $Mn$ and $V$ sublattice magnetizations at different temperatures. As a result, the ferrimagnetic
  phase is divided into two phases: in the low temperature phase $0<T<T^*$ the sublattice $Mn$ magnetization and sublattice $V$  magnetization  contribute to the magnetization of the system, while at the high temperature  $(T^*,T_N)$, the vanadium sublattice magnetization is suppressed by magnon fluctuations, and only the manganese ions have non-zero spontaneous magnetization \cite{Karchev09}. This means that high temperature phase is partial order one and $T^*$ is partial order transition temperature.
  
  The magnetic orders of vanadium and manganese electrons are antiparallel.
   Magnetic field applied, during preparation of the material, along the $Mn$ magnetic moment,
  remains $Mn$ electrons localized with saturated magnetic order, while
  $V$ electrons are delocalized. Increasing the applied magnetic field the magnetization  of $Mn$ electrons remains unchanged, as long as the compensation of the Zeeman splitting of $V$ electrons increases and respectively the magnetic moment of $V$ electrons decreases. As a result increasing the applied, during the preparation, magnetic field the total magnetization of the field cooled vanadium spinel increases below $T^*$ temperature (see Fig.\ref{(fig1)MnVexp}).

  \section {SUPERCONDUCTIVITY IN FIELD-COOLED SPIN-1/2 ANTIFERROMAGNETS\cite{Karchev17}}
  
  We discuss a novel mechanism for insulator-metal transition and superconductivity in field-cooled spin-$1/2$ antiferromagnets on bcc lattice. Applying  magnetic field on sublattice A and B electrons along the sublattice B magnetization, during preparation of the material, we change the magnetic and transport properties of the material.  Sublattice B electrons are localized, while sublattice A ones are delocalized. This remains true when applied field is switched off. {\bf Theoretically a "frozen" magnetic field should be included in the $A$ fermion dispersion, which leads effectively to decreasing of Zeeman splitting.} The effective model is a spin-fermion model with Zeeman splitting of itinerant electrons compensated by the applied field. The Hamiltonian of the system is
  \bea \label{antiferro1}\nonumber
   h   = & - & t\sum\limits_{\ll ij \gg _A } {\left( {c_{i\sigma }^ + c_{j\sigma } + h.c.} \right)}
  +U\sum\limits_{i\in A} n_{i\uparrow}n_{i\downarrow} 
   -  \mu \sum\limits_{i\in A} {n_i}
   -H \sum\limits_{i\in A} {S^{zA}_{i}} \\
  & + & J\sum\limits_{  \langle  ij  \rangle } {{\bf S_i^A}}\cdot {\bf S_j^B}
  -  J_B\sum\limits_{  \ll  ij  \gg_B  } {{\bf S_i^B}
  	\cdot {\bf S_j^B}}, 
 \eea 
where $t>0$ is the hopping parameter,  ${\bf S_i^A}$ is the spin of the itinerant
electrons at the sublattice $A$ site with components 
\be\label{antiferro1b}
S^{\nu A}_i=\frac 12\sum\limits_{\sigma\sigma'}c^+_{i\sigma}\tau^{\nu}_{\sigma\sigma'}c^{\phantom +}_{i\sigma'},\ee 
and  $(\tau^x,\tau^y,\tau^z)$ are the Pauli matrices, ${\bf S_i^B}$ is the spin of the localized electrons  at the sublattice $B$ site, $\mu$
is the chemical potential, $n_{i\sigma}=c^+_{i\sigma}c_{i\sigma}$ and $n_i=n_{i\uparrow}+n_{i\downarrow}$. The
sums are over all sites of a body centered cubic lattice, $\langle i,j\rangle$ denotes the sum over the nearest neighbors, while $ \ll  ij  \gg_A$ and  $\ll  ij  \gg_B$ are sums over all sites of sublattice $A$ and $B$ respectively. The Heisenberg term describes ferromagnetic Heisenberg exchange between sublattice B $(J_B>0)$ electrons, while the term  $J>0$ is the antiferromagnetic exchange constant between localized and itinerant electrons. The term with the constant $U>0$ is the Coulomb repulsion. The "frozen" magnetic field $H$ accounts for the effect of the applied, during the preparation, magnetic field on itinerant electrons.  
\vskip 0.5 cm
\subsection { Insulator-Metal Transition}
\vskip 0.1 cm
We represent the Fermi operators, the spin of the itinerant electrons and the density operators of sublattice A electrons in terms of the Schwinger-bosons ($\varphi_{i,\sigma}, \varphi_{i,\sigma}^+$) and slave fermions ($h_i, h_i^+,d_i,d_i^+$).  The Bose fields
are doublets $(\sigma=1,2)$ without charge, while fermions are spinless with charges 1 ($d_i$) and -1 ($h_i$).
In terms of the new fields one obtains the Hamiltonian of free fermions ($h^+,h,d^+,d$)\cite{Karchev17} 

\be\label{sup6}
h_0 =   \sum\limits_{k\in B_r} \left (\varepsilon^d_k d_k^+ d_k + \varepsilon^h_k h_k^+ h_k \right)\ee
with dispersions
\bea\label{supp7}
\varepsilon^d_k & = & -4t\varepsilon_k +U-\mu+2J-\frac H2 \nonumber \\
\varepsilon^h_k & = & 4t\varepsilon_k+\mu+2J -\frac H2  \\
\varepsilon_k &  = & \left ( \cos k_x+\cos k_y+\cos k_z \right)\nonumber \eea

The ground state of the system with Hamiltonian Eq.(\ref{sup6}), is labeled by the density of electrons
\be\label{QCB4d} n=1-<h^+_i h_i>+<d^+_id_i> \ee 
 and the zero temperature spontaneous dimensionless magnetization, of the sublattice A electron
\begin{equation}\label{QCB51}
m=\frac 12 \left(1-<h^+_i h_i>-<d^+_id_i>\right).
\end{equation}
At half-filling
\be\label{QCB4e} <h^+_i h_i>=<d^+_id_i>. \ee 
To solve this equation, for all values of the parameters $U$, $t$ and $H$, one sets the chemical potential $\mu=U/2$. 
Utilizing this representation of $\mu$ we calculate the dispersion of $"d"$ and $"h"$ fermions (\ref{supp7}) as a function of the applied magnetic field.

It is convenient to introduce the critical magnetic field
\be\label{antiferro4}
H_{cr1}=U+4J-24t.
\ee
Fermions dispersion can then be rewritten in the form
\bea\label{antiferro4a}
\varepsilon^d_k & = & 4t\left[-\varepsilon_k +3 +\frac {H_{cr1}-H}{8t} \right] \nonumber \\
\varepsilon^h_k & = &  4t\left[\varepsilon_k +3 +\frac {H_{cr1}-H}{8t} \right]
\eea
When the applied magnetic field is below the critical one $H<H_{cr1}$ the Fermion dispersions are positive ($\varepsilon^d_k>0$, \,\,$\varepsilon^h_k>0$)  for all values of the wave vector $k$. The minimum of the $d$-fermion dispersion is in the center of the Brillouin zone of a cubic lattice $B_r$ $\textbf{k}=(0,0,0)$ and $\varepsilon^d_0=(H_{cr1}-H)/2$. The minimum of the  $h$-fermion dispersion is in the vertices of the Brillouin zone $\textbf{k}^*=(\pm \pi,\pm \pi,\pm \pi)$ and $\varepsilon^h_{k^*}=(H_{cr1}-H)/2$.
This means that Fermions excitations are with gap which is our definition for insulating state. If one applies magnetic field below the critical one the prepared material is insulator.

When the applied field is above the critical one $H>H_{cr1}$, the solutions of the equations
\bea\label{antiferro4b}
\varepsilon^d_k & = & 4t\left[-\varepsilon_k +3 +\frac {H_{cr1}-H}{8t} \right]=0 \nonumber \\
\varepsilon^h_k & = &  4t\left[\varepsilon_k +3 +\frac {H_{cr1}-H}{8t} \right]=0
\eea
define the Fermi surfaces of $"d"$ and $"h"$ quasiparticles.
The resultant material is metal. The system possesses a novel insulator-metal transition when magnetic field is applied and the critical value is $H_{cr1}$ (\ref{antiferro4}).

The equation (\ref{antiferro4a}) shows that when the magnetic field is zero the system is insulator if Coulomb repulsion is strong. When a hydrostatic pressure is applied the hopping parameters $t$ increases, and for $24t>U+4J$ the system is metal. The point is that under a hydrostatic pressure all electrons in the material delocalize, while when a magnetic field is applied the electrons in the system are geometrically separated and sublattice A electrons are delocalized, but sublattice B ones are localized. This is important novelty.

\subsection { Magnon-Induced Superconductivity} 

There is a second critical value
\be\label{antiferro41}
H_{cr2}=U+4J. \ee
When $H=H_{cr2}$ the material is metal ($H_{cr2}>H_{cr1}$) and Zeeman splitting of sublattice A electrons is zero. The Fermion dispersions  $\varepsilon^d_k$ and $\varepsilon^h_k$ (\ref{antiferro4a}) adopt the form
\bea\label{antiferro4c}
\varepsilon^d_k & = & -4t\varepsilon_k  \nonumber \\
\varepsilon^h_k & = &  4t\varepsilon_k
\eea
With dispersions (\ref{antiferro4c}) spontaneous magnetization $m$ (\ref{QCB51}) of sublattice A electrons is zero and they do not contribute the magnetization of the system. At this critical point the system is in partial order state. Only sublattice B electrons are magnetically ordered, while sublattice A electrons are magnetically disordered.

When Zeeman splitting is zero we use the approximate representation for the fermi operators at quantum partial-order point (QPOP)
\be\label{QCB25}
c_{i\uparrow} = d_i, \qquad
c_{i\downarrow} = h_i^+.\ee
Then we can write the Hamiltonian of the system at QPOP in terms of the fermion operators $c_{i\sigma }^ +,\, c_{i\sigma }$ and Holstein-Primakoff (HP) bose operators $a^+_j,\,a_j$ used to represent the spin operators of sublattice B localized electrons ${\bf S^B_j}(a^+_j,a_j)$. The Hamiltonian is a sum of three terms 
\bea\label{antiferro6}
h^A & = & -t\sum\limits_{\ll ij \gg _A } {\left( {c_{i\sigma }^ + c_{j\sigma } + h.c.} \right)} \nonumber \\
h^{AB} & = & \sqrt{\frac{s}{2}}J\sum\limits_{  \langle  ij  \rangle }\left(c_{i\downarrow }^ + c_{i\uparrow }a_j+c_{i\uparrow}^ + c_{i\downarrow }a_j^+\right) \\
h^B & = &  -  J_B\sum\limits_{  \ll  ij  \gg_B  } {{\bf S_i^B}
	\cdot {\bf S_j^B}}\nonumber \eea
\vskip 0.5 cm
\ {\bf Magnon-Fermion Effective Theory}
\vskip 0.1 cm

We introduce two sublattices. 
The Hamiltonian of the  Magnon-Fermion Effective Theory, in momentum space, reads
\be\label{antiferro7}
h  =  \sum\limits_{k\in B_r}\left[ \varepsilon_{k}^A c_{k \sigma}^+ c_{k \sigma}+\varepsilon_k^B a_k^+ a_k \right] 
 +  \frac {4J\sqrt{2s}}{\sqrt{N}}\sum\limits_{k q p \in B_r } \delta (p-q-k)\cos\frac {k_x}{2} \cos\frac {k_y}{2} \cos\frac {k_z}{2} 
 \times  \left(c_{p\downarrow }^ + c_{q\uparrow }a_k+c_{q\uparrow}^ + c_{p\downarrow }a_k^+\right),\ee
with fermi $\varepsilon_k^A$ and bose $\varepsilon_{k }^B$ dispersions
\bea\label{antiferro8}
\varepsilon_{k}^A & = & -4t\left ( \cos k_x+\cos k_y+\cos k_z \right) \\
\varepsilon_k^B & = & 2J_B u \left ( 3-\cos k_x-\cos k_y-\cos k_z \right) \nonumber \eea
where the bose dispersion is calculated in Hartree-Fock approximation and $u$ is HF parameter that renormalizes the sublattice B exchange constant $J_B$.
The two equivalent sublattices A and B of the body center cubic lattice are simple cubic lattices. Therefor the wave vectors $p,q,k$ run over the first Brillouin zone of a cubic lattice $B_r$ .

Let us average in the subspace of Bosons $(a^+,a)$-to integrate the Bosons in the path integral approach. In static approximation one obtains an effective fermion theory with Hamiltonian
\be\label{antiferro9}
h_{eff}  =  \sum\limits_{k\in B_r}\varepsilon_{k}^A c_{k \sigma}^+ c_{k \sigma} - \frac 1N \sum\limits_{k_i p_i \in B_r} \delta (k_1-k_2-p_1+p_2) 
 V_{k_1-k_2} c_{k_1\downarrow}^+c_{k_2\uparrow}c_{p_2\uparrow}^+c_{p_1\downarrow} \ee
and potential
\be\label{antiferro10}
V_k = \frac {J^2 (1+\cos{k_x})(1+ \cos{k_y})(1+ \cos{k_z})}{J_B u \left ( 3-\cos k_x-\cos k_y-\cos k_z \right)}. \ee

Following standard procedure one obtains the effective Hamiltonian in the Hartree-Fock approximation
\be\label{antiferro11}
h^{HF}_{eff}=  \sum\limits_{k\in B_r}\left[ \varepsilon_{k \sigma} c_{k \sigma}^+ c_{k \sigma}+\Delta_k c_{-k\downarrow}^+c_{k\uparrow}+\Delta_k^+c_{k\uparrow}c_{-k\downarrow}\right], \ee
with gap function
\be\label{antiferro12}
\Delta_k=\frac 1N \sum\limits_{p\in B_r}<c_{-p\uparrow}c_{p\downarrow}> V_{p-k} \ee
The Hamiltonian can be written in a diagonal form by means of Bogoliubov excitations $\alpha^+,\alpha,\beta^+,\beta$ with dispersions
\be\label{antiferro13}
 E^{\alpha}_k=E^{\beta}_k=E_k = \sqrt{(\varepsilon_{k}^A)^2+|\Delta_k|^2}  \ee
In terms of the new excitations the gap equation reads
\bea\label{antiferro14}
\Delta_k= & - & \frac {1}{2N} \sum\limits_{p\in B_r}V_{k+p}\frac {\Delta_p}{\sqrt{(\varepsilon_{}^A)^2+|\Delta_p|^2}}\nonumber \\
& \times & \left(1-\frac {2}{e^{ {E_p}/{T}}+1}\right), \eea
where $T$ is the temperature.

\begin{figure}[!ht]
	\centering\includegraphics[width=4 in]{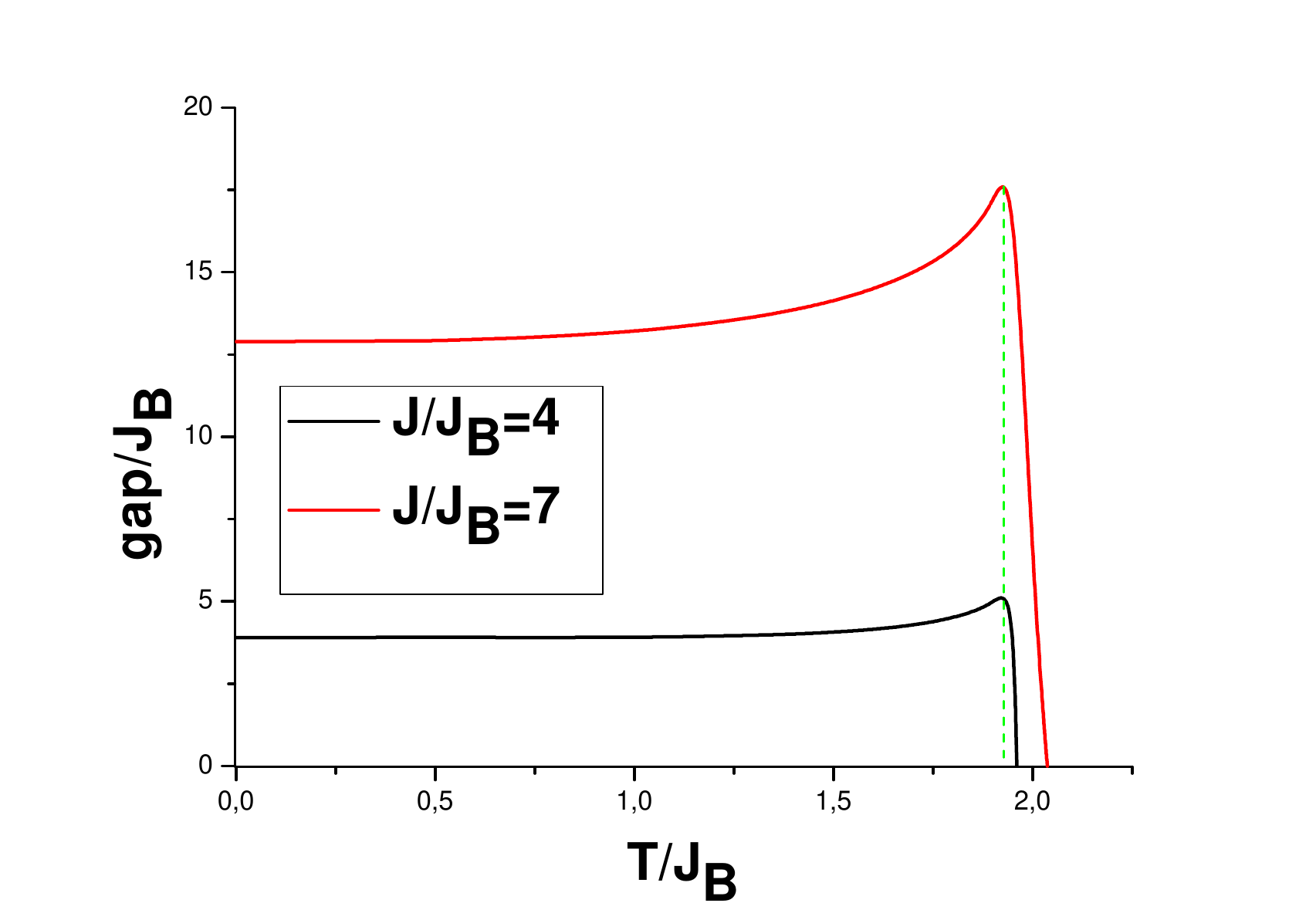}
	\epsfxsize=\linewidth
	\caption{(Color online) The temperature dependence of the dimensionless gap $(gap/J_B)$ for $t/J_B=0.5$ and two different values of the parameter $J/J_B=7$ - upper (red) graph, $J/J_B=4$ - lower (black) graph. The vertical dash (green) line marks the N\'{e}el temperature.}\label{antiferro-gap-T}
\end{figure}

Having in mind that sublattices are simple cubic lattices and following the classifications for spin-triplet gap functions $\Delta_{-k}=-\Delta_{k}$ \cite{RKS10}, we obtained that the gap function with $T_{1u}$ configuration
\be\label{antiferr15}
\Delta_k=\Delta\left(\sin k_x+\sin k_y+\sin k_z \right) \ee
is a solution of the gap equation (\ref{antiferro14}) for some values of the temperature. The dimensionless gap $gap/J_B$, as a function of dimensionless temperature $T/J_B$, is depicted in figure (\ref{antiferro-gap-T}) for two different values of the parameter $J/J_B=4$, $J/J_B=7$ and $t/J_B=0.5$.

The parameters are chosen having in mind that $J$ is nearest neighbor exchange constant while $J_B$ next to nearest  neighbor exchange constant, therefore $J>J_B$. With value $J/J_B=7$ we overestimate the parameter to give better understanding of the phenomenon.

The figure (\ref{antiferro-gap-T}) shows that the temperature dependence of the gap  is quite unusual. The gap is approximately constant when the temperature is below the N\'{e}el temperature $T_N$, marked with vertical dash green line, weakly increases when the temperature approaches $T_N$ and abruptly falls to zero in paramagnetic phase. This is because the pairing of fermions, below the N\'{e}el temperature, is mediated by gapless bosons-magnons. The potential $V_k$  depends on temperature since the  Hartree-Fock parameter $u$ does.
Near the N\'{e}el temperature the  parameter $u$ decreases \cite{Karchev17} and potential $V_k$ increases. Above N\'{e}el temperature the magnon opens a gap which rapidly increases when the temperature increases. This suppresses the superconductivity since the maximal value of the potential in paramagnetic phase is one over the magnon gap, so that when the magnon gap increases abruptly the potential decreases.

It is important to underline that the applied, during preparation of the material, magnetic field separates spatially electrons. Sublattice A ones are delocalized and participate in the formation of Cooper pairs, while B ones are localized and form the magnetic moment of the system. Thus the system possesses coexistence of p-type superconductivity and magnetism. 
At zero temperature the magnetic moment of sublattice B electrons is maximal. This is in contrast with known systems that possess coexistence of superconductivity and magnetism near quantum critical point.

{\bf The above result shows that a superconductor can be prepared from any antiferromagnet }.

\section {SEQUENCE OF SUPERCONDUCTING STATES IN FIELD COOLED $FeCr_2S_4$ \cite{Karchev21}}

In the present section we investigate two sublattice $FeCr_2S_4$ spinel. The sublattice A sites are occupied by $F^{2+}$ (s=2) iron ions, and sublattice B sites are occupied by $Cr^{3+}$ (s=3/2) chromium ions. The $Fe^{2+}$ and $Cr^{3+}$ ions are located at the center of tetrahedral and octahedral  $S^{2-}$ cages, respectively, and the three $Cr^{3+}$ electrons occupy the lower energy $t_{2g}$ bands. 

\begin{figure}[!ht]
	\centering\includegraphics[width=4in]{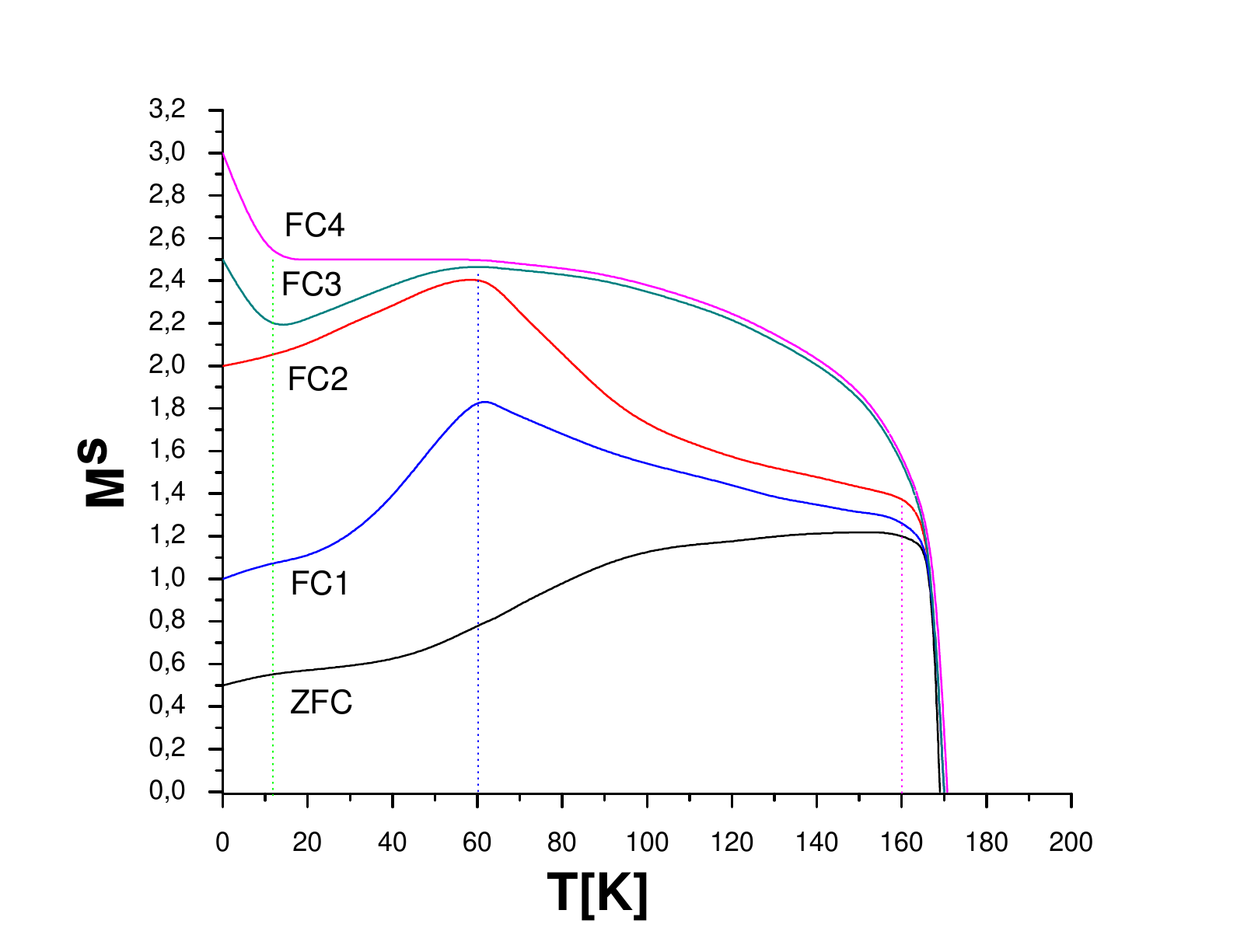}
	\epsfxsize=\linewidth
	\caption{(Color online) The temperature dependence of the spontaneous magnetization $M^s$ of ZFC, FC1, FC2, FC3 and FC4  ($FeCr_2S_4$) spinel. The ZFC, FC2 and FC3 curves are adapted from experimental results \cite{Yang00,Tsurkan10,Yto11,Lin14}. The ZFC curve has unusual maximum at $T^*=160 K $ attributed to magnetic domain dynamics in \cite{Tsurkan10}. Small dip of the curve at approximately $10K$ is interpreted as onset of long-range orbital order. It is reported \cite{JB14,Lin14} that below $50 K$ the curve  FC2 shows a transition to noncollinear  ferrimagnetism. 
		In the present paper we consider that the red, vertical line indicates the onset of magnetization order of one of the chromium electrons at temperature ($160 K$), the blue line ($60 K$) -the second electron and the green one ($10 K $) the third. The curves, FC1 and FC4, are phenomenological extrapolations based on our experience gained from the experimental and theoretical study of many other ferrimagnets.}\label{(fig6)}
\end{figure}

The shape of magnetization-temperature diagrams for all spinels is remarkable and emblematic.  In the case of $FeCr_2S_4$ spinel the curves, which show the temperature dependence of spontaneous magnetization $M^s$, are depicted in Fig.\ref{(fig6)}. 

The ZFC, FC2 and FC3 curves are adapted from experimental results \cite{Yang00,Tsurkan10,Yto11,Lin14} and the rest ones, FC1 and FC4, are phenomenological  extrapolations. 
It is important to underline the different interpretations of the behavior of the system near the four characteristic temperatures $T_N = 170 K$, $T^*=160 K$, $T=60K$ and $T=10K$ from experimental results, and those in the present paper. The N\'{e}el  temperature is $T_N = 170 K$. For ZFC spinel the temperature, at which the magnetization is maximum, is  $T^*=160 K $. 
This unusual maximum is attributed to magnetic domain dynamics in \cite{Tsurkan10}. In the same article small dip of the curve at approximately $10K$ is interpreted as onset of
long-range orbital order. 
In our interpretation supported by calculations using Green function approach \cite{Diep97} and modified spin-wave theory \cite{Karchev08}
the system undergoes a partial-order transition from the high temperature  $(T^{*},T_N)$ phase, where only the iron ions have non-zero spontaneous magnetization, to low temperature one $(0,T^{*})$, where both the iron and chromium ions have non-zero spontaneous magnetization. The exchange constant of chromium and iron spins is antiferromagnetic, because of which the magnetization increases below $T_N$ and decreases below $T^{*}$.  
The subtle point is that the onset of magnetism of the three  $Cr^{3+}$ ($t_{2g}$) electrons is at different temperatures.
One of them starts to form magnetic order below $T^{*}$. The ZFC curve in (Fig.\ref{(fig6)}), shows small dip at $10 K$ which indicates that another chromium electron starts to contribute the magnetic order of the system at this temperature. The contribution of the third  $Cr^{3+}$ electron becomes more clear in the experiment with FC spinel-curve FC2.
It shows that applied during the preparation magnetic field exceeds the Zeeman splitting energy of the chromium electron and its magnetic order becomes parallel to the iron magnetic moment. As a result, below $T^{*}$ the system undergoes partial-order transition, due to the onset of magnetic order of chromium electron and the spontaneous magnetization of the system increases and reaches the maximum at $60K$. It undergoes a second partial-order transition because of the onset of the magnetic order of the second  $Cr^{3+}$ electron anti-parallel to the iron one  and the spontaneous magnetization of the system decreases. Alternatively it is reported \cite{JB14,Lin14} that below $50 K$ the system undergoes a transition to non-collinear ferrimagnetism. Important consequence is the emergence of multiferroic phase below $10 K$. One expects that applying magnetic field, during preparation, the non-collinear order is suppressed and collinear ferrimagnetism is restored.
At $10 K$ the curve has neither dip nor increase, which means that applied magnetic field compensates Zeeman splitting of the third  chromium electron. The FC1 curve in the middle is an extrapolation for the case when the magnetic field applied during the preparation compensates the Zeeman splitting energy of the $ Cr ^ {3 +} $ electron. Therefore, in the temperature range $ (60 K-170 K) $ only iron electrons contribute to the magnetization of the system. The extrapolation is based on our experimental and theoretical knowledge acquired in the study of other spinels.
Increasing the applied, during preparation, magnetic field one obtains the magnetization-temperature curve FC3. It has two characteristic features: first the onset of the magnetism of iron and the first chromium electron is at the same temperature $T_N=170 K$, second the applied magnetic field exceeds the Zeeman splitting energy of the third chromium electron and its magnetic order becomes parallel to the iron magnetic moment. As a result, below $T=10 K$ the system undergoes partial-order transition, due to onset of the magnetic order of the  third chromium electron and the spontaneous magnetization of the system increases. The last curve FC4 is an extrapolation for the case when the applied field compensates Zeeman splitting of the second $ Cr ^ {3 +} $ electron.

Based on the analysis of Fig.\ref{(fig6)}, we consider a spin-fermion model of $FeCr_2S_4$ spinel, with three bands describing $t_{2g}$ chromium electrons and 
spin $s=2$ operators for localized $F^{2+}$ electrons. The iron-chromium exchange constants are antiferromagnetic, different for the three chromium electrons. Magnetic field in the Hamiltonian models the decrease of the Zeeman splitting during preparation of the material. We study the appearance and disappearance of superconducting states as a function of the field. The Hamiltonian of the model is
\be \label{FeCrS1}
h  =  -  t\sum\limits_{\ll ij \gg _B }\sum\limits_{\sigma,l}{\left( {c_{i\sigma l }^ + c_{j\sigma l } + h.c.} \right)}-H \sum\limits_{i\in B,\, l} {S^{zB}_{il}}
 +  \sum\limits_{  \langle  ij  \rangle, l } J_l\, {{\bf S_i^A}}\cdot {\bf S_{j l}^B}
-  J^A\sum\limits_{  \ll  ij  \gg_A  } {{\bf S_i^A}
	\cdot {\bf S_j^A}}, 
\ee 
where $S^{\nu B}_{il}=\frac 12\sum\limits_{\sigma\sigma'}c^+_{i\sigma l}\tau^{\nu}_{\sigma\sigma'}c^{\phantom +}_{i\sigma' l}$, with the Pauli
matrices $(\tau^x,\tau^y,\tau^z)$, is the spin of the $lth$ - $t_{2g}$ chromium 
electron ($l=1,2,3$) at the sublattice $B$ site , ${\bf S}_i^A$ is the spin operator of the localized iron electrons  at the sublattice $A$ site. The
sums are over all sites of a body centered cubic lattice, $\langle i,j\rangle$ denotes the sum over the nearest neighbors, $ \ll  ij  \gg_A$ is a sum over all sites of sublattice $A$ and $ \ll  ij  \gg_B$ is a sum over all sites of sublattice $B$ . The Heisenberg term $(J^A > 0)$ describes ferromagnetic
exchange between iron spins, and $J_l>0$ are the antiferromagnetic exchange constants between iron and chromium spins. $H>0$ is the "frozen" applied magnetic field in units of energy.

The compensation mechanism of the field induced superconductivity suggests that the formation of Cooper pairs is possible when Zeeman splitting of electrons is compensated by the applied magnetic field.
We choose the exchange constants well separated
$J_1<J_2<J_3$,
so that if the magnetic field $H$ compensates the Zeeman splitting of one of $t_{2g}$ chromium electrons, it is far from the compensation of Zeeman energy  of the other two electrons.  

With this in mind, we can simplify our study. When the value of the "frozen"  magnetic field is close to the Zeeman  energy of one of $t_{2g}$ chromium electrons, we can consider one band spin-fermion model of this electron instead of model (\ref{FeCrS1}). The contribution of dropped fermions can be accounted for by appropriate choice of the parameters.  In this way we 
consider three independent, one band spin-fermion models. In momentum space representation, the Hamiltonians $h_l$ ($l=1,2,3$) have the form   
\be\label{FeCrS2} \nonumber
h_l  =  \sum\limits_{k\in B_r} \varepsilon_k a_k^+ a_k  + \sum\limits_{k\in B_r \sigma} \varepsilon_{k \sigma l} c_{k \sigma l}^+ c_{k \sigma l}
 + \frac {4J_l\sqrt{2s}}{\sqrt{N}}\sum\limits_{k q p \in B_r } \delta (p-q-k)\cos\frac {k_x}{2} \cos\frac {k_y}{2} \cos\frac {k_z}{2} 
 \left(c_{p\downarrow l}^ + c_{q\uparrow l}a_k+c_{q\uparrow l}^ + c_{p\downarrow l}a_k^+\right) ,  \ee
with bose dispersion $\varepsilon_k$ of spin ($s=2$) iron magnons 
\be\label{FeCrS3}
\varepsilon_k  =  2sJ^A\left (3-\cos k_x-\cos k_y-\cos k_z\right),\ee 
and fermi $\varepsilon_{k \sigma l}$ dispersions of chromium electrons
\bea\label{FeCrS4}
\varepsilon_{k \uparrow l} & = & -2t\left ( \cos k_x+\cos k_y+\cos k_z \right)+\frac {8sJ_l-H}{2} \nonumber \\
\\
\varepsilon_{k \downarrow l} & = & -2t(\cos k_x+\cos k_y+\cos k_z)-\frac {8sJ_l-H}{2}. \nonumber
\eea    
The bosons $( a_k^+ a_k )$ are introduced by means of Holstein-Primakoff representation of the spin-2 operators of localized iron electrons.

To proceed we account for the spin fluctuations of iron, and in static approximation obtain three effective four-fermion theories with Hamiltonians $h^{eff}_l$, 
\be\label{}
h^{eff}_l = \sum\limits_{k\in B_r \sigma} \varepsilon_{k \sigma l} c_{k \sigma l}^+ c_{k \sigma l} 
 - \frac 1N \sum\limits_{k_i p_i \in B_r} \delta (k_1-k_2-p_1+p_2) 
 V_{k_1-k_2}^l c_{k_1\downarrow l}^+c_{k_2\uparrow l}c_{p_2\uparrow l}^+c_{p_1\downarrow l} \ee
and potentials
\be\label{IFer}
V_{k}^l= \frac { J^2_l  (1+\cos k_x )(1+\cos k_y )(1+ \cos k_z )}{J^A\left ( 3-\cos k_x-\cos k_y-\cos k_z \right)} \ee
The Hamiltonians in the Hartree-Fock approximation are
\be\label{}
h_l^{HF}= \sum\limits_{k\in B_r}\left[ \varepsilon_{k \sigma l} c_{k \sigma l}^+ c_{k \sigma l}+\Delta_{k l} c_{-k\downarrow l}^+c_{k\uparrow l}+\Delta_{k l}^+c_{k\uparrow l}c_{-k\downarrow l}\right],\ee
with gap functions
\be\label{}
\Delta_{k l}=\frac 1N \sum\limits_{p\in B_r}<c_{-p\uparrow l}c_{p\downarrow l}> V_{p-k }^l \ee
In terms of  Bogoliubov excitations $\alpha_l^+,\alpha_l,\beta_l^+,\beta_l$, with dispersions
\bea\label{IFerri10}
E^{\alpha}_{k l} & = & \frac 12 \left[\varepsilon_{k\uparrow l}-\varepsilon_{k\downarrow l}+\sqrt{(\varepsilon_{k\uparrow l}+\varepsilon_{k\downarrow l})^2+4|\Delta_{k l}|^2}\right] \\
E^{\beta}_{k l} & = & \frac 12 \left[-\varepsilon_{k\uparrow l}+\varepsilon_{k\downarrow l}+\sqrt{(\varepsilon_{k\uparrow l}+\varepsilon_{k\downarrow l})^2+4|\Delta_{k l}|^2}\right]. \nonumber\eea
the gap equations have the form
\bea\label{IFerri11}\nonumber
\Delta_{k l}= & - & \frac 1N \sum\limits_{p\in B_r}V_{k+p}^l\frac {\Delta_{p l}}{\sqrt{(\varepsilon_{p\uparrow l}+\varepsilon_{p\downarrow l})^2+4|\Delta_{p l}|^2}} \\
& \times & \left(1-<\alpha^+_{p l}\alpha_{p l}>-<\beta^+_{p l}\beta_{p l}>\right), \eea
where $<\alpha^+_{p l}\alpha_{p l}>$ and $<\beta^+_{p l}\beta_{p l}>$ are fermi functions for Bogoliubov fermions.

\begin{figure}[!ht]	\centering\includegraphics[width=4in]{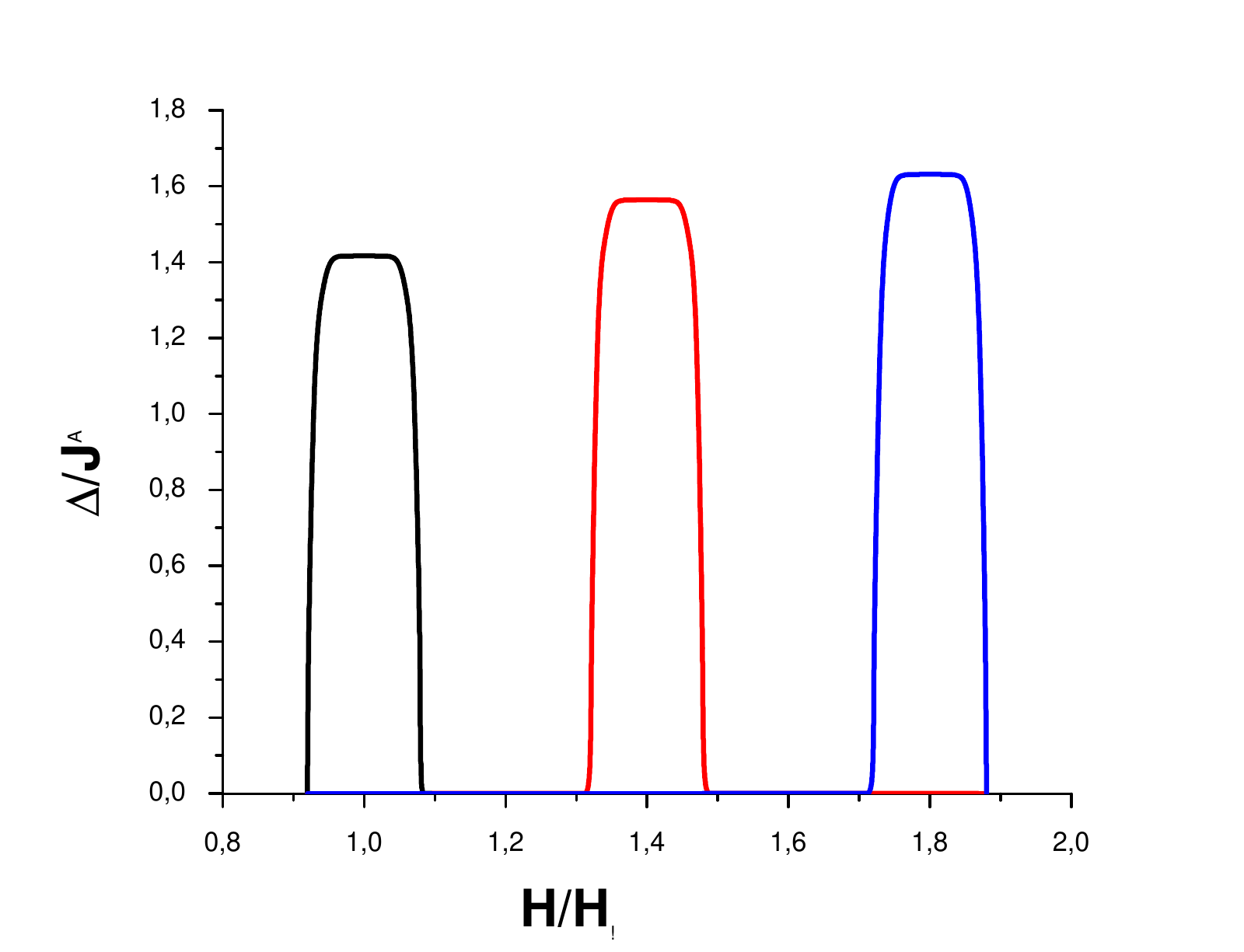}
	\epsfxsize=\linewidth
	\caption{(Color online)Sequence of superconducting states in field cooled $FeCr_2S_4$. The first state is realized near $H=H_1=2sJ_1$, the second one near $H=H_2=2sJ_2=1.4 H_1$, and the third state near $H=H_3=2sJ_3=1.8 H_1$. } \label{(fig7)}
\end{figure}

Having in mind that sublattices are simple cubic lattices and following the classifications for spin-triplet gap functions $\Delta_{-k l}=-\Delta_{k l}$, we obtained that the gap functions with $T_{1u}$ configuration \cite{RKS10}
\be\label{IFerri12} \Delta_{k l}=\Delta_l\left(\sin k_x+\sin k_y+\sin k_z) \right) \ee
are solutions of the gap equations for some values of the applied, during the preparation, magnetic field and temperature. The dimensionless gaps $\Delta_l/J^A$ at zero temperature,  as a function of $H/H_1$ where $H_1=8sJ_1$, are depicted in Fig.(\ref{(fig7)}) for parameters $J_1/J^A=2$, $J_2/J_1=1.4$, $J_3/J_1=1.8$ and $t/J^A=1$. 

The applied magnetic field $H_1$ compensates the Zeeman splitting of $t_{2g}$ chromium electrons with minimum   Zeeman energy (\ref{FeCrS4}). The Fig.(\ref{(fig7)}) shows that near this value the above mentioned electrons form Cooper pairs and superconductivity emerges. Increasing the magnetic field we restore the Zeeman splitting with opposite sign and suppress the superconductivity. 
Further increasing the magnetic field, we reach $H_2=2sJ_2$ that compensates the Zeeman splitting of another $t_{2g}$ electrons. Now, the Cooper pairs are formed by the second group electrons and the superconductivity is restored near $H/H_1 = H_2/H_1 = J_2/J_1=1.4$. This process continues until the magnetic field, applied during preparation, becomes equal to Zeeman energy of the third group of chromium electrons $H_3 =2sJ_3$ and third superconductor state emerges near $H_3/H_1=1.8$. In that way we can create a sequence of superconducting states in field cooled $FeCr_2S_4$.
Actually there are three different superconductors prepared applying, during preparation, different magnetic fields. In these compounds superconductivity coexist with the saturated magnetism of iron ions.

In summary, we have predicted the possibility to synthesize three different superconductors. The difficult moment is to applied a field which compensates the Zeeman splitting of one of the chromium electrons. An useful guidance to do that are the curves depicted in figure (\ref{(fig6)}).    
The curves FC1, FC2 and FC4 illustrate the three cases when one of chromium electrons is with compensated Zeeman splitting. They are well separated from the others which permits the exact choice of the applied, during preparation, magnetic field.

The spinel $FeCr_2S_4$ is well studied compound, but superconductivity has not been observed. The explanation is very simple. The focus of research is on colossal magnetoresistance effect (CMR) \cite{RCK1997,Yang00} and on the existence of multiferroic phase. The investigation of CMR requires measurements of resistivity as a function of temperature, but they are realized for ZFC materials only. In the rare cases, when measuring the resistivity of FC compounds, there is no clarity about the applied magnetic field. This can be achieved by simultaneous study spontaneous magnetization and resistivity as functions of temperature, which would lead to  correct choice of field. 

The experiments with spinel stimulate the study in detail some perspective antiferromagnets.
Closest to the  $FeCr_2S_4$ spinels are calcium manganese oxides ($CaMnO_3$). They are G-type antiferromagnetic insulators with N\'eel temperature $T_N=350K$. As in the case with spinels, the electrons of the $t_{2g}$ triplet contribute magnetic order. 
Experiments must be performed to show the degeneracy or non-degeneracy of the $t_{2g}$ states, which will ultimately determine how to prepare the superconductor.

\section {DO NOT CONFUSE WITH JACCARINO-PETER COMPENSATION MECHANISM OF SUPERCONDUCTIVITY \cite{JP62}}

The magnetic field induced superconductivity (FISC) is one more issue of special interest. Experimentally, (FISC) in the $H_{c2}-T$ phase diagram was observed in $Eu_xSn_{1-x}Mo_6S_8$ \cite{Fischer75}. The domain of superconductivity in $Eu_{0.75}Sn_{0.25}Mo_6S_8$ extends from 4 to 22 T at $T=0$ and from $T=0$ to $T=1 K$ at $H = 12 T$ \cite{Fischer84}. The magnetic-field induced superconductivity was attributed to Jaccarino-Peter (JP) compensation mechanism\cite{JP62}. The idea is that in a ferromagnetic metal the conduction electrons are in an effective field due to the exchange interaction with the localized spins. It is in general so large as to inhibit the occurrence of superconductivity. For some systems the exchange interaction have a negative sign. This  allows for the conduction electron polarization to be canceled by an external magnetic field so that if, in addition these metals possess phonon-induced attractive electron-electron interaction, superconductivity  occurs in the compensation region. In more complicated cases superconductivity can occur in two domains: one extends from zero applied magnetic field to small field which suppress bose condensation of Cooper pairs and respectively superconductivity, and the other at the high field in the compensation region \cite{Fischer84}.
The experiments show that the compensation  field is not affected by superconductivity.

A great deal of interest has been centered on the heavy fermions in cerium and uranium systems. The heavy-fermion system $CePb_3$ at zero field is an antiferromagnet. In \cite{Lin85} the authors report  magnetic field of 14 T induces the system into the superconducting state below 0.20 K. Similarly, at 0.48 K, 15 T magnetic field  drives the sample superconducting. 
The (FISC) in these compounds is considered to be due to the Jaccarino-Peter mechanism, extended to antiferromagnetic superconductors \cite{Shimahara02}. 

$URhGe$ displays ferromagnetism with magnetic moment oriented along the c -axis, and spin-triplet superconductivity at a lower temperature \cite{Hardy05}.
In an external magnetic field along the b-axis perpendicular to c-axis, superconductivity disappears at about $H=2 T$. However, at higher magnetic fields,
in the range from $8 T$ to $13.5 T$, it reappears again \cite{Sheikin05}.

Finally, magnetic-field-induced superconductivity has been observed in organic superconductors  \cite{Uji01,Balicas01,Konoike04}.

{\bf It is important to emphasize, that magnetic-field induced superconductivity disappears when the field is switched off.}

\end{document}